\documentclass{article}

\usepackage{arxiv}

\usepackage[utf8]{inputenc} 
\usepackage[T1]{fontenc}    
\usepackage{hyperref}       
\usepackage{url}            
\usepackage{booktabs}       
\usepackage{amsfonts}       
\usepackage{amsmath}        
\usepackage{nicefrac}       
\usepackage{microtype}      
\usepackage{cleveref}       
\usepackage{graphicx}
\usepackage{doi}
\usepackage{amsthm}         

\usepackage{algorithm}
\usepackage{algpseudocode}

\title{An O(1) Space Algorithm for N-Dimensional Tensor Rotation: A Generalization of the Reversal Method}

\date{\today}

\author{ 
    \href{https://orcid.org/0009-0008-3774-5598}{\includegraphics[scale=0.06]{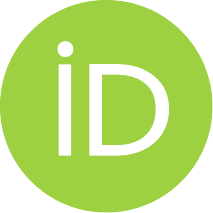}\hspace{1mm}Dexin Chen} \\
	School of Computer Science and Engineering\\
	South China University of Technology\\
	Guangzhou, China \\
	\texttt{f14xuanlv@gmail.com} \\
}

\renewcommand{\undertitle}{}
\renewcommand{\shorttitle}{Generalized Reversal Algorithm for Tensor Cyclic Shift}

\hypersetup{
pdftitle={An O(1) Space Algorithm for N-Dimensional Tensor Rotation: A Generalization of the Reversal Method},
pdfsubject={cs.DS},
pdfauthor={Your Name},
pdfkeywords={Algorithm, In-Place, Tensor Rotation, Cyclic Shift, Array Rotation, Reversal Algorithm, Multidimensional Array},
}

\begin{document}
\maketitle

\begin{abstract}
	The rotation of multi-dimensional arrays, or tensors, is a fundamental operation in computer science with applications ranging from data processing to scientific computing. While various methods exist, achieving this rotation in-place (i.e., with O(1) auxiliary space) presents a significant algorithmic challenge. The elegant three-reversal algorithm provides a well-known O(1) space solution for one-dimensional arrays.  This paper introduces a generalization of this method to N dimensions, resulting in the "$2^n+1$ reversal algorithm".  This algorithm achieves in-place tensor rotation with O(1) auxiliary space and a time complexity linear in the number of elements.  We provide a formal definition for N-dimensional tensor reversal, present the algorithm with detailed pseudocode, and offer a rigorous proof of its correctness, demonstrating that the pattern observed in one dimension ($2^1+1=3$ reversals) and two dimensions ($2^2+1=5$ reversals) holds for any arbitrary number of dimensions.
\end{abstract}


\section{Introduction}

Array rotation (cyclic shift) is a classic problem in computer science, frequently encountered in coding interviews and as a subroutine in more complex algorithms. The task is to cyclically shift the elements of an array by a given number of positions. \textbf{Terminology note:} Throughout this paper, we use the term ``rotation'' following the established convention in algorithm literature \cite{bentley1999programming}. This refers to cyclic shifting of array elements, not to be confused with geometric rotation (e.g., rotating an image by 90 degrees).  For a one-dimensional array, a straightforward approach involves using an auxiliary array, which requires O(n) additional space.

Cyclic shifting of multi-dimensional arrays is widely used in scientific computing (e.g., simulations with periodic boundary conditions), signal processing, and image processing. Popular libraries such as NumPy (\texttt{numpy.roll}) and MATLAB (\texttt{circshift}) provide convenient implementations, but these functions allocate $O(N)$ auxiliary space to store the result. For memory-constrained environments—such as embedded systems, GPU computing where device memory is limited, or large-scale scientific simulations where tensors may occupy gigabytes of memory—an in-place algorithm with $O(1)$ auxiliary space offers significant advantages. The $2^n+1$ reversal algorithm addresses this need.

A more sophisticated and space-efficient solution is the three-reversal algorithm, famously popularized by Jon Bentley in his book \emph{Programming Pearls} \cite{bentley1999programming}. This in-place method achieves rotation with O(1) auxiliary space by performing three sequential reversal operations on segments of the array. Its elegance and efficiency have made it a canonical example of algorithmic thinking.

When we extend the problem to two dimensions, the task becomes rotating a matrix where columns are shifted by $p$ units and rows are shifted by $q$ units.  Again, a naive solution would require O(m$\times$n) auxiliary space.  This naturally leads to the question: can the reversal-based approach be extended to two dimensions?  The answer is yes, and it requires five reversal operations on sub-matrices.  We observe that one dimension requires 3 reversals and two dimensions require 5 reversals, which raises an intriguing question: can the reversal method be generalized to arrays of arbitrary dimensions? Is there a fixed mathematical relationship between the number of reversals required and the dimensionality of the array?

In this paper, we answer the question of whether this pattern can be generalized to an arbitrary number of dimensions.  We affirmatively propose the \textbf{$2^n+1$ reversal algorithm}, a novel method for performing in-place, O(1) space rotation on an N-dimensional tensor. We provide a formal framework for N-dimensional rotation and reversal, present the algorithm, prove its correctness, and analyze its complexity.

\section{Definitions for N-Dimensional Tensor Rotation and Reversal}

To formalize the algorithm, we first define the concepts of N-dimensional tensor rotation and N-dimensional tensor reversal.

\subsection{N-Dimensional Tensor Rotation (Cyclic Shift)}
Let $T$ be an n-dimensional tensor of shape $(d_0, d_1, \dots, d_{n-1})$. An element in $T$ is accessed by an index vector $\mathbf{i} = (i_0, i_1, \dots, i_{n-1})$, where $0 \le i_l < d_l$ for all $l \in \{0, \dots, n-1\}$.

An N-dimensional tensor rotation is defined by a rotation vector $\mathbf{k} = (k_0, k_1, \dots, k_{n-1})$, where each $k_l$ specifies the cyclic shift along the $l$-th dimension. The rotated tensor $T'$ is defined such that the element originally at index $\mathbf{i}$ is moved to a new index $\mathbf{j}$, where:
\begin{equation}
	j_l = (i_l + k_l) \pmod{d_l} \quad \forall l = 0, \dots, n-1
\end{equation}
Equivalently, the element at index $\mathbf{j}$ in the new tensor $T'$, denoted $T'_{\mathbf{j}}$, is sourced from the element at index $\mathbf{i}$ in the original tensor $T$, where $i_l = (j_l - k_l) \pmod{d_l}$. Thus:
\begin{equation}
	T'_{\mathbf{j}} = T_{(j_0 - k_0) \pmod{d_0}, \dots, (j_{n-1} - k_{n-1}) \pmod{d_{n-1}}}
\end{equation}
For practical purposes, the rotation vector $\mathbf{k}$ is first normalized such that $k_l = k_l \pmod{d_l}$.

\textbf{Example (1D):} For a one-dimensional array $[1, 2, 3, 4, 5]$ of shape $(5)$ with cyclic shift $k_0 = 2$:
\[
\begin{bmatrix}
1 & 2 & 3 & 4 & 5
\end{bmatrix}
\xrightarrow{k_0 = 2}
\begin{bmatrix}
4 & 5 & 1 & 2 & 3
\end{bmatrix}
\]
Each element at index $i$ moves to index $(i + 2) \mod 5$.  For instance, element $1$ at index $0$ moves to index $(0+2) \mod 5 = 2$.

\textbf{Example (2D):} For a matrix of shape $(2, 3)$ with cyclic shift vector $\mathbf{k} = (1, 2)$:
\[
\begin{bmatrix}
1 & 2 & 3 \\
4 & 5 & 6 \\
\end{bmatrix}
\xrightarrow{\mathbf{k} = (1, 2)}
\begin{bmatrix}
5 & 6 & 4 \\
2 & 3 & 1 \\
\end{bmatrix}
\]
The element at index $(i_0, i_1)$ moves to index $((i_0 + 1) \mod 2, (i_1 + 2) \mod 3)$.  For instance, element $1$ at index $(0, 0)$ moves to index $(1, 2)$.

\subsection{N-Dimensional Tensor Reversal}
We generalize the one-dimensional reversal operation to n dimensions. The N-dimensional tensor reversal, denoted by $Rev$, is an operation that swaps elements symmetrically about the geometric center of the tensor (or a specified sub-tensor).

For a tensor $T$ of shape $(d_0, \dots, d_{n-1})$, the reversed tensor $T_{rev} = Rev(T)$ is defined by:
\begin{equation}
	(T_{rev})_{\mathbf{i}} = (T_{rev})_{i_0, \dots, i_{n-1}} = T_{d_0 - 1 - i_0, \dots, d_{n-1} - 1 - i_{n-1}}
\end{equation}
A crucial property of the reversal operation is that it is an \textbf{involution}. If we consider $Rev$ as an operator acting on the space of tensors, applying it twice results in the identity map $I$. Formally:
\begin{equation}
	Rev^2 = Rev \circ Rev = I
\end{equation}
This implies that for any tensor $T$:
\begin{equation}
	Rev(Rev(T)) = I(T) = T
\end{equation}
This involution property is the cornerstone of our proposed algorithm.

\textbf{Example (1D):} For a one-dimensional array of shape $(5)$:
\[
\begin{bmatrix}
1 & 2 & 3 & 4 & 5
\end{bmatrix}
\xrightarrow{Rev}
\begin{bmatrix}
5 & 4 & 3 & 2 & 1
\end{bmatrix}
\xrightarrow{Rev}
\begin{bmatrix}
1 & 2 & 3 & 4 & 5
\end{bmatrix}
\]
The element at index $i$ is swapped with the element at index $4-i$. For instance, element $2$ at index $1$ moves to index $4-1=3$.  Applying the reversal twice restores the original array.

\textbf{Example (2D):} For a matrix of shape $(2, 3)$:
\[
\begin{bmatrix}
1 & 2 & 3 \\
4 & 5 & 6 \\
\end{bmatrix}
\xrightarrow{Rev}
\begin{bmatrix}
6 & 5 & 4 \\
3 & 2 & 1 \\
\end{bmatrix}
\xrightarrow{Rev}
\begin{bmatrix}
1 & 2 & 3 \\
4 & 5 & 6 \\
\end{bmatrix}
\]
The element at index $(i_0, i_1)$ is swapped with the element at index $(1-i_0, 2-i_1)$. For instance, element $2$ at index $(0, 1)$ moves to index $(1, 1)$. Applying the reversal twice restores the original matrix.

\section[The 2\^{}n+1 Reversal Algorithm]{The $2^n+1$ Reversal Algorithm}

We now present the algorithm for performing an in-place rotation of an N-dimensional tensor $T$ given a rotation vector $\mathbf{k}$.  The high-level steps are as follows:
\begin{enumerate}
	\item \textbf{Normalization:} Normalize the rotation vector $\mathbf{k}$. 
	\item \textbf{Step 1: Global Reversal (1 reversal):} Perform a single N-dimensional reversal on the entire tensor $T$.
	\item \textbf{Step 2: Partitioning:} The rotation vector $\mathbf{k}$ conceptually partitions the tensor into $2^n$ hyperrectangular sub-tensors (blocks). 
	\item \textbf{Step 3: Block-wise Reversals ($2^n$ reversals):} Perform an N-dimensional reversal on each of the $2^n$ sub-tensors. 
\end{enumerate}
The detailed logic is presented in Algorithm \ref{alg:reverse} and Algorithm \ref{alg:rotate}.

\begin{algorithm}[H]
\caption{In-Place N-Dimensional Sub-Tensor Reversal}
\label{alg:reverse}
\begin{algorithmic}[1]
\State \textbf{function} ReverseND(Tensor $T$, StartIndices $\mathbf{s}$, EndIndices $\mathbf{e}$)
\State \quad $n \gets \text{dimensionality of } T$
\State \quad \text{Initialize current index } $\mathbf{i} \gets \mathbf{s}$
\State \quad \textbf{loop}
\State \quad \quad \text{Calculate mirror index } $\mathbf{j} \gets \mathbf{s} + \mathbf{e} - \mathbf{i}$ \Comment{Component-wise operation}
\State \quad \quad \textbf{if} $\mathbf{i}$ is lexicographically $\ge$ $\mathbf{j}$ \textbf{then}
\State \quad \quad \quad \textbf{break} \Comment{All pairs have been swapped}
\State \quad \quad \textbf{end if}
\State \quad \quad swap($T[\mathbf{i}], T[\mathbf{j}]$)
\State \quad \quad \Comment{Increment index $\mathbf{i}$ to the next position}
\State \quad \quad $d \gets n - 1$
\State \quad \quad \textbf{while} $d \ge 0$ \textbf{do}
\State \quad \quad \quad \textbf{if} $i_d < e_d$ \textbf{then}
\State \quad \quad \quad \quad $i_d \gets i_d + 1$
\State \quad \quad \quad \quad \textbf{break}
\State \quad \quad \quad \textbf{else}
\State \quad \quad \quad \quad $i_d \gets s_d$
\State \quad \quad \quad \quad $d \gets d - 1$
\State \quad \quad \quad \textbf{end if}
\State \quad \quad \textbf{end while}
\State \quad \quad \textbf{if} $d < 0$ \textbf{then break} \Comment{All indices exhausted}
\State \quad \textbf{end loop}
\State \textbf{end function}
\end{algorithmic}
\end{algorithm}

\begin{algorithm}[H]
\caption{The $2^n+1$ Reversal Algorithm for N-Dimensional Tensor Rotation}
\label{alg:rotate}
\begin{algorithmic}[1]
\State \textbf{function} RotateND(Tensor $T$, RotationVector $\mathbf{k}$)
\State \quad $n \gets \text{dimensionality of } T$
\State \quad $\mathbf{d} \gets \text{shape of } T$
\State \quad \Comment{Step 0: Normalize rotation vector}
\State \quad \textbf{for} $l \gets 0$ \textbf{to} $n-1$ \textbf{do}
\State \quad \quad $k_l \gets k_l \pmod{d_l}$
\State \quad \textbf{end for}
\State
\State \quad \Comment{Step 1: Global Reversal}
\State \quad $\mathbf{s}_{global} \gets (0, \dots, 0)$; \quad $\mathbf{e}_{global} \gets (d_0-1, \dots, d_{n-1}-1)$
\State \quad ReverseND($T, \mathbf{s}_{global}, \mathbf{e}_{global}$)
\State
\State \quad \Comment{Steps 2 \& 3: Partition and perform $2^n$ block-wise reversals}
\State \quad \textbf{for} $i \gets 0$ \textbf{to} $2^n - 1$ \textbf{do}
\State \quad \quad \text{Initialize block boundaries } $\mathbf{s}_{block}, \mathbf{e}_{block}$
\State \quad \quad \textbf{for} $l \gets 0$ \textbf{to} $n-1$ \textbf{do}
\State \quad \quad \quad \textbf{if} \text{the $l$-th bit of $i$ is 0} \textbf{then}
\State \quad \quad \quad \quad $s_{block, l} \gets 0$; \quad $e_{block, l} \gets k_l - 1$
\State \quad \quad \quad \textbf{else}
\State \quad \quad \quad \quad $s_{block, l} \gets k_l$; \quad $e_{block, l} \gets d_l - 1$
\State \quad \quad \quad \textbf{end if}
\State \quad \quad \textbf{end for}
\State \quad \quad \textbf{if} $\forall l: s_{block, l} \le e_{block, l}$ \textbf{then} \Comment{Check for valid block}
\State \quad \quad \quad ReverseND($T, \mathbf{s}_{block}, \mathbf{e}_{block}$)
\State \quad \quad \textbf{end if}
\State \quad \textbf{end for}
\State \textbf{end function}
\end{algorithmic}
\end{algorithm}

\section{Worked Examples}

To build intuition for the algorithm before presenting the formal proof, we trace the algorithm for the one-dimensional and two-dimensional cases.  Each case is presented first with abstract notation, then with concrete numerical values.

\subsection{The 1D Case: \texorpdfstring{$2^1+1=3$}{2\^{}1+1=3} Reversals}

\subsubsection{Abstract Representation}

Consider a one-dimensional array of length $d$ with cyclic shift $k$ (assumed to be normalized, i.e., $0 \le k < d$).  The array is partitioned into two segments:
\[
[\underbrace{A}_{[0, d-k-1]} \mid \underbrace{B}_{[d-k, d-1]}]
\]
where $A$ denotes the first $d-k$ elements and $B$ denotes the last $k$ elements.

The goal of cyclic shift by $k$ positions to the right is to transform $[A \mid B]$ into $[B \mid A]$.

\begin{enumerate}
    \item \textbf{Step 1 -- Global Reversal:}
    \[
    [A \mid B]
    \xrightarrow{Rev}
    [Rev(B) \mid Rev(A)]
    \]
    
    \item \textbf{Step 2 -- Block-wise Reversals:}
    \[
    [Rev(B) \mid Rev(A)]
    \xrightarrow{Rev_1, Rev_2}
    [Rev(Rev(B)) \mid Rev(Rev(A))]
    = [B \mid A]
    \]
\end{enumerate}

The cyclic shift is achieved using $2^1+1=3$ reversal operations.

\subsubsection{Numerical Example}

Consider the array $[1, 2, 3, 4, 5, 6, 7]$ of shape $(7)$ with cyclic shift $k = 3$:
\[
\left[\begin{array}{cccc|ccc}
1 & 2 & 3 & 4 & 5 & 6 & 7 \\
\end{array}\right]
\]
where $A = [1,2,3,4]$ and $B = [5,6,7]$.

\begin{enumerate}
    \item \textbf{Step 1 -- Global Reversal:}
    \[
    \left[\begin{array}{cccc|ccc}
    1 & 2 & 3 & 4 & 5 & 6 & 7 \\
    \end{array}\right]
    \xrightarrow{Rev}
    \left[\begin{array}{ccc|cccc}
    7 & 6 & 5 & 4 & 3 & 2 & 1 \\
    \end{array}\right]
    \]
    
    \item \textbf{Step 2 -- Block-wise Reversals:}
    \[
    \left[\begin{array}{ccc|cccc}
    7 & 6 & 5 & 4 & 3 & 2 & 1 \\
    \end{array}\right]
    \xrightarrow{Rev_1, Rev_2}
    \left[\begin{array}{ccc|cccc}
    5 & 6 & 7 & 1 & 2 & 3 & 4 \\
    \end{array}\right]
    \]
\end{enumerate}

The final result $[5, 6, 7, 1, 2, 3, 4]$ achieves the desired cyclic shift by $k=3$ positions to the right.

\subsection{The 2D Case: \texorpdfstring{$2^2+1=5$}{2\^{}2+1=5} Reversals}

\subsubsection{Abstract Representation}

Consider a two-dimensional matrix with cyclic shift vector $\mathbf{k}=(k_0, k_1)$ (assumed to be normalized, i.e., $0 \le k_0 < d_0$ and $0 \le k_1 < d_1$). The matrix is partitioned into four blocks ($2^2=4$) by the binary vector $\mathbf{b}$:

\[
\begin{bmatrix}
B_{(0,0)} & B_{(0,1)} \\
B_{(1,0)} & B_{(1,1)} \\
\end{bmatrix}
\]
where:
\begin{itemize}
    \item $B_{(0,0)}$: indices $(0,0)$ to $(k_0-1, k_1-1)$ (first $k_0$ rows, first $k_1$ columns)
    \item $B_{(0,1)}$: indices $(0,k_1)$ to $(k_0-1, d_1-1)$ (first $k_0$ rows, last $d_1-k_1$ columns)
    \item $B_{(1,0)}$: indices $(k_0,0)$ to $(d_0-1, k_1-1)$ (last $d_0-k_0$ rows, first $k_1$ columns)
    \item $B_{(1,1)}$: indices $(k_0,k_1)$ to $(d_0-1, d_1-1)$ (last $d_0-k_0$ rows, last $d_1-k_1$ columns)
\end{itemize}

The goal of cyclic shift by $(k_0, k_1)$ is to move each block $B_{\mathbf{b}}$ to the position of $B_{\bar{\mathbf{b}}}$ (its diagonally opposite block). That is, the target matrix has blocks arranged as:
\[
\begin{bmatrix}
B_{(1,1)} & B_{(1,0)} \\
B_{(0,1)} & B_{(0,0)} \\
\end{bmatrix}
\]

\begin{enumerate}
    \item \textbf{Step 1 -- Global Reversal:}
    \[
    \begin{bmatrix}
    B_{(0,0)} & B_{(0,1)} \\
    B_{(1,0)} & B_{(1,1)} \\
    \end{bmatrix}
    \xrightarrow{Rev}
    \begin{bmatrix}
    Rev(B_{(1,1)}) & Rev(B_{(1,0)}) \\
    Rev(B_{(0,1)}) & Rev(B_{(0,0)}) \\
    \end{bmatrix}
    \]
    
    \item \textbf{Step 2 -- Block-wise Reversals (4 reversals):}
    \[
    \begin{bmatrix}
    Rev(B_{(1,1)}) & Rev(B_{(1,0)}) \\
    Rev(B_{(0,1)}) & Rev(B_{(0,0)}) \\
    \end{bmatrix}
    \xrightarrow{\substack{Rev_{(0,0)}, Rev_{(0,1)}, \\ Rev_{(1,0)}, Rev_{(1,1)}}}
    \begin{bmatrix}
    B_{(1,1)} & B_{(1,0)} \\
    B_{(0,1)} & B_{(0,0)} \\
    \end{bmatrix}
    \]
\end{enumerate}

The cyclic shift is achieved using $2^2+1=5$ reversal operations.

\subsubsection{Numerical Example}

Consider the matrix of shape $(5, 7)$ with cyclic shift vector $\mathbf{k} = (2, 3)$:
\[
\left[\begin{array}{cccc|ccc}
1 & 2 & 3 & 4 & 5 & 6 & 7 \\
8 & 9 & 10 & 11 & 12 & 13 & 14 \\
15 & 16 & 17 & 18 & 19 & 20 & 21 \\
\hline
22 & 23 & 24 & 25 & 26 & 27 & 28 \\
29 & 30 & 31 & 32 & 33 & 34 & 35 \\
\end{array}\right]
\]
where:
\[
B_{(0,0)} = \begin{bmatrix} 1 & 2 & 3 & 4 \\ 8 & 9 & 10 & 11 \\ 15 & 16 & 17 & 18 \end{bmatrix},
B_{(0,1)} = \begin{bmatrix} 5 & 6 & 7 \\ 12 & 13 & 14 \\ 19 & 20 & 21 \end{bmatrix},
B_{(1,0)} = \begin{bmatrix} 22 & 23 & 24 & 25 \\ 29 & 30 & 31 & 32 \end{bmatrix},
B_{(1,1)} = \begin{bmatrix} 26 & 27 & 28 \\ 33 & 34 & 35 \end{bmatrix}
\]

\begin{enumerate}
    \item \textbf{Step 1 -- Global Reversal:}
    \[
    \left[\begin{array}{cccc|ccc}
    1 & 2 & 3 & 4 & 5 & 6 & 7 \\
    8 & 9 & 10 & 11 & 12 & 13 & 14 \\
    15 & 16 & 17 & 18 & 19 & 20 & 21 \\
    \hline
    22 & 23 & 24 & 25 & 26 & 27 & 28 \\
    29 & 30 & 31 & 32 & 33 & 34 & 35 \\
    \end{array}\right]
    \xrightarrow{Rev}
    \left[\begin{array}{ccc|cccc}
    35 & 34 & 33 & 32 & 31 & 30 & 29 \\
    28 & 27 & 26 & 25 & 24 & 23 & 22 \\
    \hline
    21 & 20 & 19 & 18 & 17 & 16 & 15 \\
    14 & 13 & 12 & 11 & 10 & 9 & 8 \\
    7 & 6 & 5 & 4 & 3 & 2 & 1 \\
    \end{array}\right]
    \]
    
    \item \textbf{Step 2 -- Block-wise Reversals:}
    \[
    \left[\begin{array}{ccc|cccc}
    35 & 34 & 33 & 32 & 31 & 30 & 29 \\
    28 & 27 & 26 & 25 & 24 & 23 & 22 \\
    \hline
    21 & 20 & 19 & 18 & 17 & 16 & 15 \\
    14 & 13 & 12 & 11 & 10 & 9 & 8 \\
    7 & 6 & 5 & 4 & 3 & 2 & 1 \\
    \end{array}\right]
    \xrightarrow{\substack{Rev_{(0,0)}, Rev_{(0,1)}, \\ Rev_{(1,0)}, Rev_{(1,1)}}}
    \left[\begin{array}{ccc|cccc}
    26 & 27 & 28 & 22 & 23 & 24 & 25 \\
    33 & 34 & 35 & 29 & 30 & 31 & 32 \\
    \hline
    5 & 6 & 7 & 1 & 2 & 3 & 4 \\
    12 & 13 & 14 & 8 & 9 & 10 & 11 \\
    19 & 20 & 21 & 15 & 16 & 17 & 18 \\
    \end{array}\right]
    \]
\end{enumerate}

The final matrix achieves the desired cyclic shift: each element originally at index $(i_0, i_1)$ is now at index $((i_0 + 2) \mod 5, (i_1 + 3) \mod 7)$.  For instance, element $1$ at index $(0, 0)$ has moved to index $(2, 3)$.

\section{Proof of Correctness}

Let $T$ be the original tensor and $T'$ be the target rotated tensor. Let $\mathbf{k}$ be the normalized rotation vector. The tensor $T$ is partitioned into $2^n$ blocks.  Let us denote the set of elements (the content) originally located in the region defined by the binary vector $\mathbf{b}$ as $B_{\mathbf{b}}$. 

The state of the tensor at a region $\mathbf{b}$ can be described as $\{ \text{content at region } \mathbf{b} \}$. The initial state is $T \cong \{ B_{\mathbf{b}} \text{ at region } \mathbf{b} \}_{\forall \mathbf{b}}$. 

The target state after rotation is $T' \cong \{ B_{\mathbf{\bar{b}}} \text{ at region } \mathbf{b} \}_{\forall \mathbf{b}}$, where $\mathbf{\bar{b}}$ is the bitwise complement of $\mathbf{b}$ (i.e., $\bar{b}_l = 1 - b_l$).  To see why this is the target state, consider that a rotation by $\mathbf{k}$ shifts elements such that an element originally in block $\mathbf{b}$ (where $b_l = 0$ means index $< k_l$, and $b_l = 1$ means index $\ge k_l$) will, after adding $\mathbf{k}$ modulo $\mathbf{d}$, end up in the block $\bar{\mathbf{b}}$.  This is because indices in $[0, k_l)$ shift to $[k_l, d_l)$, and indices in $[k_l, d_l)$ wrap around to $[0, k_l)$.  In simpler terms, the goal is to move the content from each block to its diametrically opposite block.

Let's trace the algorithm:
\begin{enumerate}
	\item \textbf{After Global Reversal:} Let $T_1 = Rev(T)$.  The global reversal operation swaps the content of region $\mathbf{b}$ with the content of region $\mathbf{\bar{b}}$, while also reversing the content within each block.  Therefore, the content now located at region $\mathbf{b}$ is the reversed content from the original region $\mathbf{\bar{b}}$, which is $Rev(B_{\mathbf{\bar{b}}})$.
	\begin{equation}
		T_1 \cong \{ Rev(B_{\mathbf{\bar{b}}}) \text{ at region } \mathbf{b} \}_{\forall \mathbf{b}}
	\end{equation}
	
	\item \textbf{After Block-wise Reversals:} The algorithm proceeds to apply a reversal operation to the content currently residing in each region $\mathbf{b}$. The content at region $\mathbf{b}$ is $Rev(B_{\mathbf{\bar{b}}})$.  Applying the reversal gives:
	\begin{equation}
		Rev(Rev(B_{\mathbf{\bar{b}}})) = Rev^2(B_{\mathbf{\bar{b}}}) = I(B_{\mathbf{\bar{b}}}) = B_{\mathbf{\bar{b}}}
	\end{equation}
	This utilizes the involution property ($Rev^2=I$) established in Section 2.2.
	
	\item \textbf{Final State:} After performing this operation for all $2^n$ blocks, the final tensor $T_{final}$ has the content $B_{\mathbf{\bar{b}}}$ located at region $\mathbf{b}$. 
	\begin{equation}
		T_{final} \cong \{ B_{\mathbf{\bar{b}}} \text{ at region } \mathbf{b} \}_{\forall \mathbf{b}}
	\end{equation}
\end{enumerate}
This final state is identical to the target state $T'$.  Thus, the $2^n+1$ reversal algorithm correctly performs the N-dimensional rotation. 

\section{Complexity Analysis}

\paragraph{Time Complexity. }
An in-place reversal of a sub-tensor with $N_{sub}$ elements takes $O(N_{sub})$ time, as each element is involved in at most one swap.  The global reversal takes $O(\prod d_l)$ time.  The sum of the sizes of all $2^n$ blocks is equal to the size of the entire tensor.  Therefore, the total time for all block-wise reversals is also $O(\prod d_l)$.  The overall time complexity is dominated by these reversal operations, resulting in a time complexity of $O(\prod_{l=0}^{n-1} d_l)$, which is linear in the number of elements in the tensor.

\paragraph{Space Complexity.}
The auxiliary space required is $O(n)$ for storing index vectors and loop counters, where $n$ is the number of dimensions. Since $n$ is typically a small constant independent of the data size $N = \prod_{l=0}^{n-1} d_l$, this is considered $O(1)$ auxiliary space with respect to the input data size.

\section{Parallelization Analysis}

The $2^n+1$ reversal algorithm is highly parallelizable. We analyze its parallel structure and derive the parallel time complexity.

\subsection{Parallel Structure}

The parallelism in this algorithm can be understood at two levels:

\begin{enumerate}
    \item \textbf{Intra-block parallelism:} Within any reversal operation on $N_{sub}$ elements, all $\lfloor N_{sub}/2 \rfloor$ swap pairs $(\mathbf{i}, \mathbf{j})$ access disjoint memory locations, as each index appears in exactly one pair. Therefore, all swaps within a single reversal can execute simultaneously.
    
    \item \textbf{Inter-block parallelism:} In Stage 2, the $2^n$ blocks are mutually disjoint (i.e., $R_{\mathbf{b}_1} \cap R_{\mathbf{b}_2} = \emptyset$ for $\mathbf{b}_1 \neq \mathbf{b}_2$), so their reversals can execute concurrently.
\end{enumerate}

Notably, if intra-block parallelism is fully exploited (i.e., all swaps within each reversal execute simultaneously), inter-block parallelism is achieved automatically. This is because the swaps from different blocks also access disjoint memory locations. In other words:
\begin{equation}
    \text{Intra-block parallelism} \Rightarrow \text{Inter-block parallelism (implicitly)}
\end{equation}

Thus, from a unified perspective, the entire algorithm reduces to two fully parallel stages:
\begin{itemize}
    \item \textbf{Stage 1:} $\lfloor N/2 \rfloor$ independent swaps (global reversal)
    \item \textbf{Stage 2:} $\sum_{\mathbf{b}} \lfloor N_{\mathbf{b}}/2 \rfloor \le \lfloor N/2 \rfloor$ independent swaps (all block reversals combined)
\end{itemize}

\subsection{Parallel Time Complexity}

Let $N = \prod_{l=0}^{n-1} d_l$ be the total number of elements and $P$ be the number of processors.

\begin{table}[H]
\centering
\caption{Sequential vs. parallel complexity}
\label{tab:parallel}
\begin{tabular}{@{}lcc@{}}
\toprule
\textbf{Metric} & \textbf{Sequential} & \textbf{Parallel ($P$ processors)} \\
\midrule
Time & $O(N)$ & $O(N/P)$ \\
Work & $O(N)$ & $O(N)$ \\
Speedup & --- & $\Theta(P)$ \\
Space & $O(n)$ & $O(n \cdot P)$ \\
\bottomrule
\end{tabular}
\end{table}

The algorithm achieves near-optimal linear speedup. The maximum useful parallelism is $P_{max} = \lfloor N/2 \rfloor$, beyond which adding more processors yields no further improvement.

\section{Conclusion}

We have presented the $2^n+1$ reversal algorithm, a novel and efficient method for the in-place rotation of N-dimensional tensors.  By generalizing the well-known three-reversal technique for one-dimensional arrays, we have shown that a similar, elegant solution exists for any arbitrary number of dimensions.  The algorithm's correctness is established through a formal proof based on the involutory property of the N-dimensional reversal operation.  With its O(1) auxiliary space complexity (with respect to data size), this algorithm provides a practical and memory-efficient solution for a fundamental data manipulation problem. 

Furthermore, the algorithm exhibits excellent parallelization potential due to the independence of swap operations within each reversal and the disjointness of blocks, achieving near-optimal $\Theta(P)$ speedup with $P$ processors.

Future work could involve implementing and benchmarking this algorithm in high-performance computing libraries for applications in image processing and scientific simulations, as well as exploring its performance characteristics based on memory layouts such as row-major and column-major order. 

\bibliographystyle{unsrtnat}

\end{document}